\documentclass[floats,floatfix,amssymb,prd,twocolumn,superscriptaddress,nofootinbib]{revtex4-2}

\usepackage[english]{babel}
\usepackage[utf8]{inputenc}
\usepackage{amsmath}
\usepackage{mathbbol}
\usepackage{amssymb}
\usepackage{bbold}
\usepackage{graphicx,amsfonts}
\usepackage{epsfig}
\usepackage[svgnames]{xcolor}
\definecolor{crimsonglory}{rgb}{0.75,0.0,0.2}
\usepackage[colorlinks=true,
linkcolor=crimsonglory,
urlcolor=crimsonglory,
citecolor=crimsonglory]{hyperref}
\usepackage{bm}
\usepackage{mathrsfs}
\usepackage{mathtools}
\usepackage{enumerate}
\usepackage{amsthm}
\usepackage{bbm}
\usepackage{comment}
\usepackage{physics}
\usepackage{url}
\usepackage{upgreek}
\usepackage[title]{appendix}
\usepackage{float} 
\usepackage{rotating}
\usepackage{booktabs}

\definecolor{maroon}{cmyk}{0,0.87,0.68,0.32}

\begin{document}

\title{Perturbing the vortex: quasinormal and quasibound spectra of rotating acoustic geometries}

\author{H. S. Vieira}
\affiliation{Theoretical Astrophysics, Institute for Astronomy and Astrophysics, University of T\"{u}bingen, 72076 T\"{u}bingen, Germany}

\author{Kyriakos Destounis}
\affiliation{Theoretical Astrophysics, Institute for Astronomy and Astrophysics, University of T\"{u}bingen, 72076 T\"{u}bingen, Germany}
\affiliation{CENTRA, Departamento de Física, Instituto Superior Técnico – IST, Universidade
de Lisboa – UL, Avenida Rovisco Pais 1, 1049-001 Lisboa, Portugal}

\author{Kostas D. Kokkotas}
\affiliation{Theoretical Astrophysics, Institute for Astronomy and Astrophysics, University of T\"{u}bingen, 72076 T\"{u}bingen, Germany}

\begin{abstract}
Strong-field gravity simulators are laboratory experiments that can investigate a wide range of both classical and quantum phenomena occurring in nature. In this work, we introduce an effective geometry that captures most of the characteristics of the strong-field regime of astrophysical, rotating black holes. This geometry can represent a vortex made from a variety of fluid and superfluid profiles with zero viscosity, making it a promising finite-temperature quantum-field-theory simulator for rotating curved spacetimes. Our geometry includes not only the typical radial flow which gives rise to an acoustic horizon, but also azimuthal circulation of the fluid. We compute the quasinormal modes, semi-analytically, and the exact quasibound states of acoustic excitations interacting with this effective geometry. The resulting spectra can be identified for both co-rotating and counter-rotating surface acoustic waves. In particular, the behavior of our acoustic geometry with circulation aligns with the phenomenology observed in recent experiments that include superfluids.
\end{abstract}

\maketitle

%%%%%%%%%%%%%%%%%%%%%%%%%%%%%%%%%%%%%%%%%%
\section{Introduction}\label{Introduction}
%%%%%%%%%%%%%%%%%%%%%%%%%%%%%%%%%%%%%%%%%%

Analog black-hole (BH) experiments \cite{Visser:1997ux,Fischer:2001jz,Barcelo:2005fc} are currently playing an important role towards a more complete understanding of classical and quantum effects around astrophysical BHs \cite{Barack:2018yly}. From such gravity simulators, we have so far obtained information regarding BH spectroscopy \cite{Kokkotas:1999bd,Nollert:1999ji,Berti:2009kk,Konoplya:2011qq,Schutzhold:2002rf,Dolan:2011ti,Torres:2020tzs,Berti:2004ju,Cardoso:2004fi,Torres:2019sbr,Patrick:2018orp,Vieira:2021ozg,Vieira:2019dau,Vieira:2021xqw,Vieira:2023ylz,Destounis:2023ruj,Rosato:2024arw,Pezzella:2024tkf,Jaramillo:2020tuu,Destounis:2021lum,Boyanov:2022ark,Destounis:2023nmb,Boyanov:2023qqf,Cheung:2021bol,Courty:2023rxk,Boyanov:2024fgc}, superradiance \cite{1972JETP...35.1085Z,Brito:2015oca,Patrick:2020baa,Torres:2016iee,Richartz:2014lda,Richartz:2012bd,Mascher:2022pku,Mollicone:2024lxy}, Hawking radiation \cite{Brout:1995wp,Giovanazzi:2004zv,Giovanazzi:2006nd,Weinfurtner:2010nu,Belgiorno:2010wn,Steinhauer:2014dra,Coutant:2016vsf}, and BH evaporation \cite{Unruh:1980cg,Jacobson:1991gr,Unruh:1994je}. These research topics have either recently begun to emerge through gravitational-wave (GW) detections \cite{LIGO:2021ppb,KAGRA:2021vkt} or have not yet been explored in fully relativistic astrophysical setups
\cite{Barausse:2020rsu,LISA:2022kgy,LISA:2022yao,Karnesis:2022vdp,Cardoso:2021wlq}. Ultimately, we expect that the critical prognosis provided by various analog gravity simulations, will further elucidate the classical, semi-classical \cite{2023PhRvD.108l4053A,2024PhRvD.109i6014A, 2024PhRvD.110f4084A} and quantum nature of spacetime \cite{Hawking:1974rv,Hawking:1975vcx}.

Unruh's revolutionary idea on testing, experimentally, Hawking radiation and BH evaporation \cite{Unruh:1980cg,Unruh:1994je} through gravitational analogs gave rise to a profusion of cutting-edge experiments. The original idea of Unruh was based on  perturbation-inducing waves on the surface of a fluid, i.e. \emph{acoustic waves}. In his setup, sound excitations can be mapped to the equations of motion of perturbations (wave equations) which are central in BH dynamics. Fluctuations on the surface of the fluid create an effective geometry characterized by
the propagation speed of these fluctuations and their relative speed with respect to the fluid, leading to an \emph{acoustic analog BH}. The radial velocity of the fluid acts as a tuning parameter of radial flow into a draining vortex. When the inflow becomes supersonic an \emph{acoustic horizon} forms, beyond which sound perturbations cannot escape the vortex. This is the definition of an analog BH that can be constructed in the laboratory. 

Currently, a multitude of gravity simulators have been built using various fluids \cite{Euve:2015vml,Euve:2017gfj,Rousseaux:2007is,Weinfurtner:2010nu,Richartz:2014lda,Cardoso:2016zvz,Torres:2016iee,Patrick:2018orp,Patrick:2019kis,Torres:2019sbr,Torres:2020tzs,Euve:2021mnj,Ciszak:2021xlw,Jannes:2010sa,Jacobson:1998ms}, Bose-Einstein condensates \cite{Garay:2000jj,Klaers:2010,Klaers:2010_2,Gooding:2020scc,Liao:2018avv,Greveling:2018,Lahav:2009wx,Fabbri:2020unn,Solnyshkov:2018dgq}, lasers \cite{Corley:1998rk,Coutant:2009cu,Leonhardt:2008js,Steinhauer:2014dra,Reyes:2017,Peloquin:2015rnl}, optical media \cite{Drori:2018ivu,Ornigotti:2017yqw,Richartz:2012bd,Philbin:2007ji,Rosenberg:2020jde}, and electric circuits \cite{Tokusumi:2018wii,Katayama:2021}, to name a few. Even though the above experiments have reproduced both realistic and theoretical phenomenology of BHs and compact objects, all experimental setups can experience instrumental limitations (water tank size, fluid dissipation \cite{Jacobson:1998ms}, bubble-forming instability at high rotation rates \cite{Andersen2003AnatomyOA}, thermal equilibrium of semiconductor lasers \cite{Schofield:2024}, only classical BH effects for optical experiments \cite{Rosenberg:2020jde}, negative-refraction metamaterials \cite{Katayama:2021,metamaterials}, dispersion \cite{Svancara:2023yrf}). Thus, the more experiments performed with different instrumentation, the better we will understand the effectiveness of BH analogs. For example, the idea of using different states of matter, such as superconductivity \cite{Jacobson:1998ms,Shi:2021nkx,Katayama:2022qmr} and superfluidity \cite{Jacobson:1998ms,Svancara:2023yrf}, Bose-Einstein condensates \cite{Baak:2022hum,Ribeiro:2021fpk,Pal:2024qno}, and even ferromagnetism \cite{Asenjo:2011mv}, enables a more in-depth perception of analog systems, due to their efficient experimental properties. One very effective example of a state of matter that directly connects to experiments is superfluidity. This is a characteristic state of matter of a fluid when it undergoes a phase transition at low-temperatures and leads to a material that has zero viscosity and, therefore, flows without any loss of kinetic energy. Thus, when stirred, a vortex forms that rotate indefinitely which can be used to study quantum effects related to event horizons, superradiance and ergoregions \cite{Jacobson:1998ms,Jacquet:2022vak,Alperin:2020aeq,Inui:2020,Braidotti:2021nhw}, and even the dynamics of the early Universe and cosmology \cite{Krusius:1998ms,Volovik:2003fe,Fischer:2004bf,Fedichev:2003bv,Fedichev:2003id}. 

In this work, we employ Unruh's fluid flow acoustic metric and include \emph{circulation}, i.e. a nonzero angular flow velocity, to obtain a mimicker of a spinning BH near the acoustic horizon, defined by a radial velocity flow (as in \cite{Berti:2004ju}). We are interested in the new aspects of the spinning metric and especially its full spectrum against small excitations. In the following sections, we first build the acoustic BH with circulation, in order to devise a targeted correspondence with contemporary experiments. We find the quasinormal modes (QNMs) and quasibound states (QBSs) of such effective geometry and present the results throughout the whole parameter space of the metric. We find the exact spectrum of QBSs by the use of confluent Heun functions, and observe that they have a generic resemblance with current experiments, such as the ones in Ref. \cite{Patrick:2018orp,Svancara:2023yrf}. Regarding QNMs, we embark on a novel numerical analysis with the current metric by including new parameters in the BH analog. Due to their wide radial profile, QNMs in experiments are sometimes hard to identify, a phenomenon that does not occur for QBSs that have narrow oscillation amplitudes. This analysis, together with the state-of-the-art \cite{Torres:2019sbr,Torres:2020tzs,Matyjasek:2024uwo} on analog BH spectroscopy, may further assist the experimental community in the detection of QNMs and support efforts to enhance the BH spectroscopy program \cite{Destounis:2023ruj,Ringdown_review} for fluids and superfluids.

%%%%%%%%%%%%%%%%%%%%%%%%%%%%%%%%%%%%%%%%%%%%%%
\section{Effective acoustic black holes with circulation}\label{UABHc}
%%%%%%%%%%%%%%%%%%%%%%%%%%%%%%%%%%%%%%%%%%%%%%

We start by examining a general acoustic BH geometry in Minkowski spacetime as derived by Unruh \cite{Unruh:1980cg} (see also Refs.~\cite{Visser:1997ux,Barcelo:2005fc}). The choice of the radial component of the fluid-flow velocity is discussed. The novelty in our geometry is the inclusion of circulation in the background to simulate the rotational flow of the vortex in the azimuthal direction induced by some experimental apparatus\footnote{Other attempts to simulate rotating draining bathtub vortices have been made in \cite{Berti:2004ju}, where QNMs and superradiance were analyzed}. We consider a suitable choice for the angular part of the flow velocity, and hence we obtain an acoustic metric that assumes the form of a rotating BH near the acoustic horizon. The fundamental equations of motion for an \emph{irrotational} fluid are given by
%%%%%%%%%
\begin{eqnarray}
\nabla \times \mathbf{v} & = & 0 \, , \label{eq:irrotational}\\
\partial_{t}\rho+\nabla\cdot(\rho\mathbf{v}) & = & 0 \,, \label{eq:Continuity}\\
\rho[\partial_{t}\mathbf{v}+(\mathbf{v}\cdot\nabla)\mathbf{v}] +\nabla p & = &0 \, , \label{eq:Euler}
\end{eqnarray}
%%%%%%%%%
where $\mathbf{v}$, $\rho$, and $p$ are the velocity, density, and pressure of the fluid, respectively. We introduce the velocity potential $\Psi$, such that $\mathbf{v}=-\nabla\Psi$, and assume the fluid as barotropic, which means that $\rho=\rho(p)$. Then, by linearizing these equations of motion around some background $(\rho_{0},p_{0},\Psi_{0})$, namely,
\begin{eqnarray}
\rho & = & \rho_{0}+\epsilon\rho_{1}, \label{eq:rho}\\
p & = & p_{0}+\epsilon p_{1}, \label{eq:p}\\
\Psi & = & \Psi_{0}+\epsilon\Psi_{1}, \label{eq:psi}
\label{eq2:Madelung_representation}
\end{eqnarray}
we obtain the wave equation
\begin{align}
&-\partial_{t}\biggl[c_s^{-2}\rho_{0}(\partial_{t}\Psi_{1}+\mathbf{v}_{0}\cdot\nabla\Psi_{1})\biggr]\nonumber\\
&+\nabla\cdot\biggl[\rho_{0}\nabla\Psi_{1}-c_s^{-2} \rho_{0}\mathbf{v}_{0}(\partial_{t}\Psi_{1}+\mathbf{v}_{0}\cdot\nabla\Psi_{1})\biggr]=0,
\label{eq:wave_equation_Visser}
\end{align}
%%%%%
where $c_{s}$ is the local speed of sound defined by
%%%%%%%
\begin{equation}
c_{s}^{-2} \equiv \frac{\partial \rho}{\partial p}\, .
\label{eq:sound}
\end{equation}
%%%%%%%
The wave equation (\ref{eq:wave_equation_Visser}) describes the propagation of the linearized scalar potential $\Psi_{1}$, that is, it governs the propagation of the phase fluctuations as weak excitations in a homogeneous stationary fluid, which can be rewritten as a wave equation in curved spacetime
%%%%%%
\begin{equation}
\frac{1}{\sqrt{-g}}\partial_{\mu}(g^{\mu\nu}\sqrt{-g}\partial_{\nu}\Psi_{1})=0.
\label{eq:KG_equation}
\end{equation}
%%%%%%
Note that this wave equation is similar to the covariant massless Klein-Gordon equation, where the corresponding acoustic line element can be written as
%%%%%%
\begin{equation}
ds^{2} = \frac{\rho_{0}}{c_{s}}\biggl[-c_{s}^{2}dt^{2}+(dx^{i}-v_{0}^{i}dt)\delta_{ij}(dx^{j}-v_{0}^{j}dt)\biggr].
\label{eq:acoustic_metric}
\end{equation}
%%%%%
Unruh assumed the background flow as a spherically-symmetric, stationary inviscid fluid. By including circulation, i.e., rotation, to the radial flow, the acoustic metric given by Eq.~(\ref{eq:acoustic_metric}) becomes
%%%%%%
\begin{align}
\nonumber
ds^{2}&=\frac{\rho_{0}}{c_{s}}\biggl\{-\biggl[c_{s}^{2}-(v_{0}^{r})^{2}-(v_{0}^{\phi})^{2}\biggr]dt^{2}\\
\nonumber
&+\frac{c_{s}}{c_{s}^{2}-(v_{0}^{r})^{2}}dr^{2}-2v_{0}^{\phi}r\sin\theta dtd\phi\\
&+r^{2}\left(d\theta^{2}+\sin^{2}\theta d\phi^{2}\right)\biggr\},
\label{eq:Unruh_circulation}
\end{align}
%%%%%%%%
where the following coordinate transformations were performed
%%%%%%%%%
\begin{eqnarray}
&& dt \rightarrow dt-\frac{v_{0}^{r}}{c_{s}^{2}-(v_{0}^{r})^{2}}dr,\\
&& d\phi \rightarrow d\phi-\frac{v_{0}^{\phi}v_{0}^{r}}{r\sin\theta[c_{s}^{2}-(v_{0}^{r})^{2}]}dr,
\label{eq:coordinate_transformations}
\end{eqnarray}
%%%%%%%%%
with 
$\mathbf{ v}=v_{0}^{r}\mathbf{e}_r+v_{0}^{\phi}\mathbf{e}_\phi$. 
Since the geometry is rotating, the term $c_{s}^{2}-(v_{0}^{r})^{2}-(v_{0}^{\phi})^{2}=0$ provides a condition for the ergosphere, which in turn is connected to the extraction of energy from rotating BHs and the eventual phenomenon of superradiance \cite{1972JETP...35.1085Z,Torres:2016iee,Brito:2015oca}.

If the background radial flow smoothly exceeds the supersonic velocity at $r=r_{h}$, then $r_h$ is an acoustic horizon and the radial component of the radial flow velocity can be expanded there as
%%%%%%%%%
\begin{equation}
v_{0}^{r}=-c_{s}+a(r-r_{h})+\mathcal{O}(r-r_{h})^{2},
\label{eq:radial_velocity}
\end{equation}
%%%%%%%%%
where the tuning parameter $a$ is defined as
%%%%%%%%%
\begin{equation}
a=(\nabla\cdot\mathbf{v})|_{r=r_{h}}.
\label{eq:a_4DUABH}
\end{equation}
%%%%%%%%%
The angular component of the flow velocity can be written as
%%%%%%%%%
\begin{equation}
v_{0}^{\phi}=\frac{J}{r}=\frac{C\sin\theta}{r},
\label{eq:angular_velocity}
\end{equation}
%%%%%%%%%
where $C$ and $J$ can be identified as the circulation and angular momentum of the fluid, respectively. Therefore, we can write the line element for an acoustic BH with circulation as
%%%%%%%%%
\begin{align}\nonumber
ds^{2}&=-\biggl[f(r)-\frac{J^{2}}{r^{2}}\biggr]dt^{2}+\frac{1}{f(r)}dr^{2}\\
&-2J\sin\theta dtd\phi+r^{2}(d\theta^{2}+\sin^{2}\theta d\phi^{2})\, ,
\label{eq:metric_UABHc}
\end{align}
%%%%%%%%%
where the acoustic metric function $f(r)$ takes the form
%%%%%%%%%
\begin{equation}
f(r)=2a(r-r_{h}),
\label{eq:metric_function_UABHc}
\end{equation}
%%%%%%%%%
with the acoustic event horizon $r_{h}$ being the outermost marginally trapped surface for outgoing phonons, where the radial velocity of the fluid becomes supersonic. Note that we have dropped a position-independent prefactor by fixing $c_{s}=1$ and assumed a constant fluid density $\rho_{0}$. The ergosphere in Eq. \eqref{eq:metric_UABHc} can be found by solving 
%%%%%%%%%
\begin{equation}
2a(r-r_h)=\frac{J^2}{r^2},  
\end{equation}
%%%%%%%%%
which gives a unique real solution, i.e.,
%%%%%%%%%
\begin{equation}
r_\textrm{erg} = \frac{y^{1/3}}{6}+\frac{2}{3y^{1/3}}+\frac{1}{3},
\end{equation}
%%%%%%%%%
where $y \equiv 8r_h^3 + 108k + 12\sqrt{81k^2 + 12kr_h^3}$, and $k \equiv J^2/2a$. In addition, the acoustic surface gravity (or the gravitational acceleration) on the background acoustic horizon surface is given by
%%%%%%%%%
\begin{equation}
\kappa_{h} \equiv \frac{1}{2}\frac{df(r)}{dr}\biggr|_{r=r_{h}} = a.
\label{eq:surface gravity_UABHc}
\end{equation}
%%%%%%%%%
Furthermore, the dragging angular velocity of the acoustic horizon, $\Omega$, is given by
%%%%%%%%%
\begin{equation}
\Omega(r) \equiv -\frac{g_{03}}{g_{33}} = \frac{C}{r^{2}},
\label{eq:dragging_UABHc}
\end{equation}
%%%%%%%%%
such that $\Omega_{h}\equiv \Omega(r_{h})=C/r_{h}^{2}$ is the frame dragging at the acoustic event horizon $r=r_{h}$.

The geometry obtained here, in Eq. \eqref{eq:metric_UABHc}, always has an acoustic horizon at $r=r_h$, therefore it is analogous to a simplified rotating BH. One of the experiments in the literature, namely the one in Ref. \cite{Svancara:2023yrf}, achieves two different horizonless states; one that does not have a depthless hollow and another one that forms a depthless hollow vortex. The second case is more akin to BH spacetimes and therefore, our effective metric in Eq. \eqref{eq:metric_UABHc} could, in principle, simulate a depthless hollow vortex of a rotating (super)fluid, though without non-linear dispersive terms that appear in Ref. \cite{Svancara:2023yrf} and further complicate the setup.

In what follows, we will analyze the motion of acoustic excitations propagating at the external region of the acoustic vortex spacetime \eqref{eq:metric_UABHc} and calculate their quasinormal and quasibound spectra. We will adopt the sixth-order Wentzel-Kramers-Brillouin (WKB) approximation to compute the QNMs, i.e. the oscillations that dominate the BH ringdown, and the Vieira-Bezerra-Kokkotas (VBK) approach to obtain the \emph{exact spectrum} of QBSs, which are bound states confined between the effective potential of the acoustic geometry and infinity.

%%%%%%%%%%%%%%%%%%%%%%%%%%%%%%%%%%%%%%%%
\section{Scalar wave equation}\label{WE}
%%%%%%%%%%%%%%%%%%%%%%%%%%%%%%%%%%%%%%%%

We are interested in the basic features of our effective geometry in Eq. \eqref{eq:metric_UABHc}, in particular the propagation of excitations in its surface, i.e. the QNM and the QBS spectra. In order to perform this analysis, we have to solve the wave equation (\ref{eq:KG_equation}) by first applying appropriate boundary conditions for each type of spectra. Since the line element \eqref{eq:metric_UABHc} resembles those of slowly-rotating spacetimes (even though $g_{tt}$ depends on both $r$ and $\theta$) we consider a spherically-symmetric-like separation ansatz for the perturbation $\Psi_1$, i.e.,
\begin{equation}
\Psi_{1}(t,r,\theta,\phi)\simeq\mbox{e}^{-i \omega t}U(r)P_{\nu}^{m}(\cos\theta)\mbox{e}^{i m \phi},
\label{eq:ansatz_UABHc}
\end{equation}
where $\omega$ is the frequency, $U(r)=R(r)/r$ is the radial function, and $P_{\nu}^{m}(\cos\theta)$ are the associated Legendre functions with general degree $\nu$ $(\in \mathbb{C})$ and order $m \geq 0$ $(\in \mathbb{Z})$, such that $\nu$ and $m$ are the angular and magnetic quantum numbers, respectively. In the case of QNMs, $\nu$ is a real number and corresponds to the real angular number $\ell$, known in the literature, while for the case of QBSs, $\nu$ is a complex number, as we will discuss in what follows. By substituting Eqs.~(\ref{eq:angular_velocity})-(\ref{eq:ansatz_UABHc}) into Eq.~(\ref{eq:KG_equation}), we obtain the radial master equation given by
%%%%%
\begin{align}
\nonumber
&R''(r)+\frac{f'(r)}{f(r)}R'(r)\\
&\quad +\frac{r^{2}\left[\omega-m\Omega(r)\right]^{2}-f(r)\left[\lambda+rf'(r)\right]}{r^{2}f^{2}(r)}R(r)=0,
\label{eq:radial_equation_UABHc}
\end{align}
where $\lambda=\nu(\nu+1)$ is the separation constant. From Eq.~(\ref{eq:radial_equation_UABHc}), we can see that, for superradiance to occur, the oscillation frequency of the perturbation $\omega$ must be less than a critical value $\omega_{c}(r)=m\,\Omega(r)$ \cite{Berti:2004ju}.

\begin{figure*}[t]
	\centering
    \includegraphics[width=0.99\textwidth]{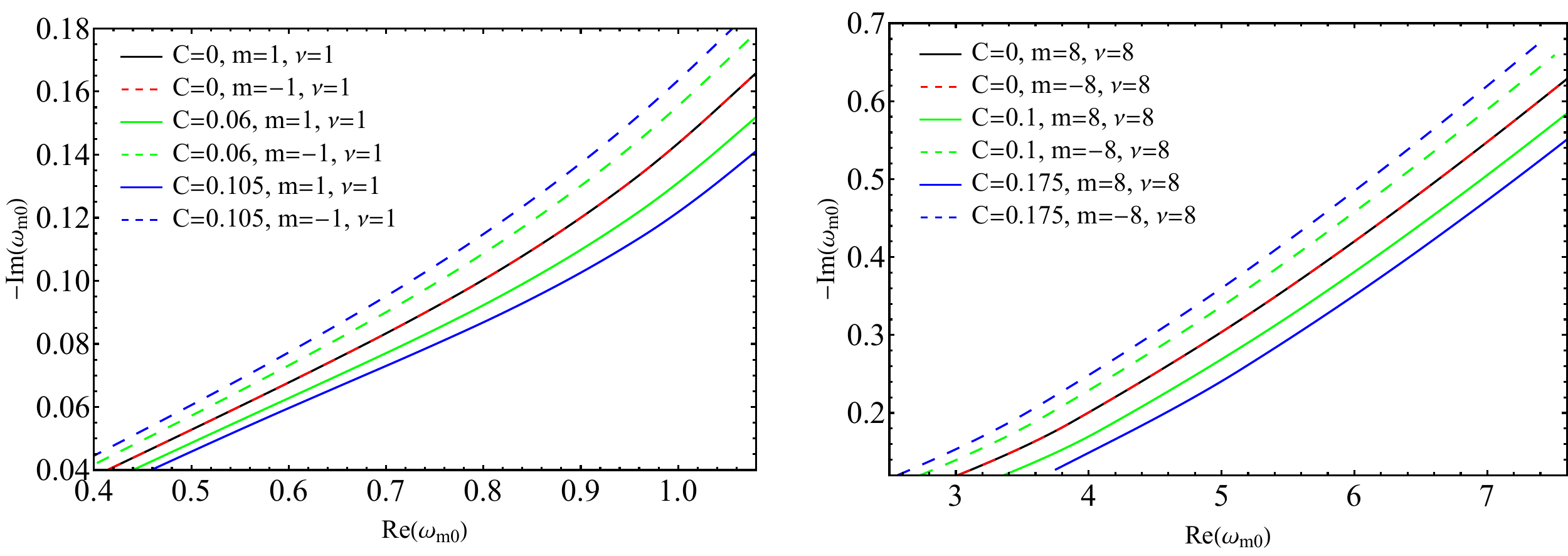}
	\caption{\emph{Left:} Corotating ($m=1$) and counterrotating ($m=-1$) $\nu=1$ fundamental QNMs, for varying tuning parameter $a=0.1,\dots, 0.45$, with step $0.05$, and various circulation parameters $C$. \emph{Right:} Corotating ($m=8$) and counterrotating ($m=-8$) $\nu=8$ fundamental QNMs, for varying tuning parameter $a=0.1,\dots, 1.5$, with step $0.2$, and various circulation parameters $C$. We note that for $C=0$ the solid black and the dashed red lines overlap.}
	\label{fig:Fig1_QNMs_UABHc}
\end{figure*}

%%%%%%%%%%%%%%%%%%%%%%%%%%%%%%%%%%%%%%%%%%
\subsection{Quasinormal modes}\label{QNMs}
%%%%%%%%%%%%%%%%%%%%%%%%%%%%%%%%%%%%%%%%%%

By multiplying Eq. \eqref{eq:radial_equation_UABHc} with $f^2(r)$ we obtain a Schr\"odinger-like equation of the form
\begin{equation}\label{eq:QNM_master}
    \frac{d^2 R(r)}{dr_*^2}+(\omega^2-\mathcal{V}) R(r)=0,
\end{equation}
where the effective potential $\mathcal{V}$ is
\begin{equation}
    \mathcal{V}\equiv f(r)\biggl[\frac{\lambda}{r^2}+\frac{f{'} (r)}{r}\biggr]-\frac{C^2 m^2}{r^4}+\frac{2Cm \omega}{r^2},
    \label{eq:potential}
\end{equation}
and $dr/dr_*=f(r)$ is the tortoise coordinate. To find the QNMs one has to solve Eq. \eqref{eq:QNM_master} with QNM boundary conditions, that is, purely ingoing excitations at the acoustic horizon and purely outgoing excitations at infinity. Note that the difference between QNM and QBS boundary conditions is at infinity, where QBSs decay there.

The effect of circulation is clearly imprinted in the effective potential in Eq. \eqref{eq:potential}. This effect makes the potential $\omega$-dependent. However, there are still an abundance of analytical \cite{Ferrari:1984zz,Cardoso:2008bp,Destounis:2018utr}, semi-analytical \cite{Schutz:1985km,Iyer:1986np,Iyer:1986nq,Kokkotas:1988fm,Seidel:1989bp,Kokkotas:1991vz,Kokkotas:1993ef,Konoplya:2002ky,Konoplya:2003ii,Matyjasek:2017psv,Konoplya:2019hlu}, and advanced numerical methods \cite{1975RSPSA.344..441C,1985RSPSA.402..285L,Gundlach:1993tp,Pani:2013pma,Jansen:2017oag,Lin:2016sch,Lin:2017oag,Lin:2019mmf,Shen:2022ssv,Lin:2023rkd,Cardoso:2017soq,Cardoso:2018nvb,Destounis:2018qnb,Liu:2019lon,Destounis:2019hca,Destounis:2019omd,Destounis:2020pjk,Destounis:2020yav,Destounis:2022rpk} to calculate QNMs. Due to the problem in hand, we will use the sixth-order WKB approximation, adapted properly so that it can function with $\omega$-dependent effective potentials \cite{Konoplya:2002ky,Vieira:2021ozg}. The procedure is practically the same as for $\omega$-independent potentials. The difference that arises is that the position $r$ at which the $\mathcal{V}(r,\omega)$ in Eq. \eqref{eq:potential} is maximized is found as a numerical function of $\omega$. The remaining procedure is identical for any WKB order approximation. Despite the fact that the WKB approximation does not converge with the increment of order expansion at the peak of the potential, we applied the same methodology used in \cite{Kokkotas:1991vz,Kokkotas:1993ef} to find the QNMs of Kerr and Kerr-Newman BHs.

The first-order WKB approximation for QNMs with respect to $a$ (assuming $r_h=1$, $m=0$ and $\nu=2$) can be found analytically as
\begin{equation}
    \omega^2 =\frac{a(a+3)}{3} \left[ (a+3) \pm i\frac{2\sqrt{3a}}{3} (3-a) \left(n+\frac{1}{2}\right)\right],
    \label{eq:solut1}
\end{equation}
where $n=0,\,1,\,2\,\dots$ is the overtone number. In what follows we will calculate and present the QNMs of the effective geometry \eqref{eq:metric_UABHc}.

The behavior of corotating and counterrotating QNMs is illustrated in Fig.~\ref{fig:Fig1_QNMs_UABHc} as a function of the tuning parameter $a$ and the circulation $C$. A universal trend is found for various values of  $C$, where both the real and the absolute value of the imaginary part of QNMs increase with the tuning parameter $a$. A symmetry around $C=0$ is observed for corotating and counterrotating modes. This is merely due to the role that $C$ plays in the potential \eqref{eq:potential}. When $C$ is set to zero, all terms that are multiplied with it, such as $m$, play no role in the dynamics of the potential, no matter the magnetic number. The QNMs provided here can, in principle, shed more light in experiments such as the one in \cite{Svancara:2023yrf}, where evidence of BH ringdown counterrotating modes ($m<0$) have been observed in hollow-core superfluid $^4$He vortices with circulation.

%%%%%%%%%%%%%%%%%%%%%%%%%%%%%%%%%%%%%%%%%%
\subsection{Quasibound states}\label{QBSs}
%%%%%%%%%%%%%%%%%%%%%%%%%%%%%%%%%%%%%%%%%%

In this section we apply the VBK approach to transform the radial equation (\ref{eq:radial_equation_UABHc}) into a confluent Heun equation without any boundary condition assumptions. Then, we will impose the QBS boundary conditions on this exact analytical radial solution, i.e., ingoing waves at the acoustic horizon and vanishing solutions at infinity, in order to obtain and analyze the QBS spectrum of the rotating acoustic geometry\footnote{A detailed discussion on the VBK approach is given in Refs.~\cite{Vieira:2021ozg,Vieira:2022pxd,Vieira:2021nha}.}.

The VBK approach essentially obtains the confluent Heun equation 
\begin{align}
\nonumber
y''(x)&+\biggl(2A_{2}+\frac{1+2A_{0}}{x}+\frac{1+2A_{1}}{x-1}\biggr)y'(x)\\
&+\biggl(\frac{A_{3}}{x}+\frac{A_{4}}{x-1}\biggr)y(x)=0,
\label{eq:CHE_UABHc}
\end{align}
where $x=1-r_h/r$  and the coefficients $A_0$, $A_1$, $A_2$, $A_3$ and $A_4$ are functions of $\omega$, $a$, $C$ and $m$ and are given in Appendix \ref{AppA}.

There are two linearly independent analytic solutions of the covariant Klein-Gordon equation at the boundaries of the effective geometry \eqref{eq:metric_UABHc}, with respect to the confluent Heun functions. By finding the asymptotic solutions at the horizon and at infinity, with imposed QBS boundary conditions, we match on the overlap region of validity by using the polynomial conditions for the Heun functions. Then, we can find the QBS spectrum. Using the resulting conditions for QBSs we find their exact spectrum as
%%%%%%%%%%%%%%%%
\begin{eqnarray}
\omega_{mn}
&=& m\Omega_h\biggl[1+\frac{a^2}{ m^2\Omega_h^2+a^2(n+1)^2}\biggr]  \nonumber \\
&&- i a (n+1)\biggl[1-\frac{a^2}{m^2\Omega_h^2 + a^2(n+1)^2}\biggr],
\label{eq:omega_UABHc1}
\end{eqnarray}
%%%%%%%%%%%%%%%%
where $n$ is the overtone number and $\Omega_h$ is the angular frequency of the acoustic horizon. Note that the QBS frequencies satisfy the symmetry $\omega_{mn}=-[\omega_{-mn}]^*$, where ``*'' denotes complex conjugation. This symmetry indicates that these solutions share the same decay rates (imaginary parts) and opposite oscillation frequency signs (real parts), which implies a symmetry under the transformation $m \rightarrow -m$. For that reason, we only show results for corotating QBSs. For the full proof, see Appendix \ref{AppA}.

\begin{figure}[t]
	\centering
	\includegraphics[width=1\columnwidth]{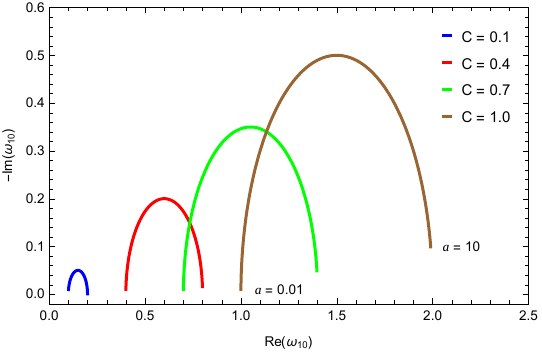}
	\caption{The fundamental $n=0$ QBSs with unitary acoustic event horizon $r_{h}=1$, varying tuning parameter $a=0.01,\dots,10$ and different choices of circulation $C$, for corotating $m=1$ modes.}
	\label{fig:Fig1_QBSs_UABHc}
\end{figure}

Even more interestingly, in the large $m$ or $n$ limit, the  QBSs will be approximately
\begin{equation}
\omega_{m n}\approx m\Omega_h -ia(n+1)=\frac{mC}{r_h^2}-ia(n+1)\, .
\label{eq:omega_UABHc1a}
\end{equation}
Alternatively, Eq. \eqref{eq:omega_UABHc1} can be written as 
%%%
\begin{eqnarray}
{\tilde \omega}_{mn}
&=& \biggl[1+\frac{{\tilde a}^2}{ 1+{\tilde a}^2(n+1)^2}\biggr]  \nonumber \\
&&- i {\tilde a} (n+1)\biggl[1-\frac{{\tilde a}^2}{1 + {\tilde a}^2(n+1)^2}\biggr]\, ,
\label{eq:omega_UABHc2}
\end{eqnarray}
%%%%%
where ${\tilde a}=a/m\Omega_h$ and ${\tilde \omega}=\omega/m\Omega_h$. Furthermore, Eq. \eqref{eq:omega_UABHc2} can be written in a more compact form as:
%%%
\begin{equation}
{\tilde \omega}_{mn}
= \left(1+{\cal A}_{mn}\right) - i {\tilde a} (n+1)\left(1-{\cal A}_{mn}\right),
\label{eq:omega_UABHc3}
\end{equation}
%%%%%
where 
%%%%
\begin{equation}
{\cal A}_{mn} = \frac{{\tilde a}^2}{1 + {\tilde a}^2(n+1)^2}.
\end{equation} 
%%%%%
In Fig. \ref{fig:Fig1_QBSs_UABHc} we demonstrate, for completeness, the corotating ($m>0$) QBSs, with a fixed acoustic horizon $r_h=1$. For counterrotating ($m<0$) QBSs, we know that they are symmetric with respect to the imaginary part, as described above, practically mirroring each other in the complex plane where the real part just changes sign. The symmetry $\mbox{Re}(\omega_{m0}) \rightarrow -\mbox{Re}(\omega_{-m0})$ is satisfied for all circulations $C$ and ranges of the tuning parameter $a$ according to the exact solution. 

Further theoretical predictions can be extracted from the exact form of QBSs. The minimum frequency permissible for propagation of QBSs for a given acoustic horizon $r_h$, $a$ and overtone $n$ is given approximately as $\sim m\Omega_h$, which is obvious from Eq. \eqref{eq:omega_UABHc1}. This means that the minimum frequency can be found by setting $r_h$ large or $C$ large, since $\Omega_h=C/r_h^2$. Obviously, the minimum frequency for $m=0$ is $0$. The same holds for large $m$. That is, the maximum frequency is determined by $m$, $C$ and $r_h$, since according to Eq. \eqref{eq:omega_UABHc1a}, the large-frequency limit is approximately $m\Omega_h$. As an example, by using the general exact spectrum in Eq. \eqref{eq:omega_UABHc1}, the fundamental QBS frequency (real part) for $a=1$, $r_h=1$, $m=10$ and $C=0.5$ is $\sim 5.192$. Using the asymptotic solution in Eq. \eqref{eq:omega_UABHc1a} we get exactly $5$, which is already quite close to the asymptotic limit $m\rightarrow\infty$. Thus the larger the $m$ the more accurately the maximum frequency can be approximated.

In conclusion, Eq. \eqref{eq:omega_UABHc1} indicates that the QBSs for varying acoustic horizons will form ``parallel'' hyperbolas as $C$ increases. We also observe that the imaginary part depends solely on the modified tuning parameter $\tilde{a}$. In fact, this equation appears to reveal a much more significant pattern of the QBS frequencies with respect to the radius of the rotating acoustic BH $r_h$, with a fixed tuning parameter $a=1$. The hyperbolic shapes of QBS frequencies for different circulation parameters $C$ become even more pronounced as $m$  increases, leading to a larger acoustic horizon. 

%%%%%%%%%%%%%%%%%%%%%%%%%%%%%%%%%%%%%%%%%%
\section{Final remarks}\label{Conclusions}
%%%%%%%%%%%%%%%%%%%%%%%%%%%%%%%%%%%%%%%%%%

Analog gravity represents the epitome of strong-gravity tests conducted in the laboratory, featuring a large number of experimental devices operating on different fundamental principles, such as water, light, lasers, condensates, circuits, and more. Undertaking these experiments provides a pathway to interpret quantum effects near black holes and classical processes that remain elusive to our detectors. In this work, we develop an effective geometry that describes a (super)fluid with circulation, which closely resembles the spacetime of a rotating black hole. The analysis of surface acoustic waves leads to semi-analytical quasi-normal modes (QNM) and the exact quasi-bound state (QBS) spectra of this geometry. Our results qualitatively align with those found in previous experimental studies, such as \cite{Patrick:2018orp, Torres:2020tzs, Svancara:2023yrf}.

Our QNM analysis, of course, yields a set of modes for different azimuthal numbers, overtones, and the parameters of the geometry — i.e., the tuning parameter  $a$  and the circulation  $C$ — without any specific assumption regarding the relationship between these two free parameters. It also reinforces the existence of ringdown modes in both corotating and counterrotating surface excitations, thereby paving the way for further experimentation in laboratory rotating vortices. Similarly, the QBS analysis seems to support some of the findings of Ref. \cite{Svancara:2023yrf}, albeit in a different, yet potentially promising, manner. The exact QBS spectra are provided for both co- and counterrotating acoustic waves in our analog rotating black hole geometry.

The future perspectives in analog gravity and the corresponding experiments are vast and can clarify the phenomenological effects of BH physics, both at the quantum and classical levels. For example, a vortex experiment might be able to test the detection of spectral instability of QNMs \cite{Cheung:2021bol,Courty:2023rxk,Destounis:2023ruj}, where a small disturbance to the vortex may lead to disproportionate migration of QNMs in the complex plane, forming logarithmic branches  \cite{Destounis:2021lum}. Another extremely interesting direction is to mimic environmental effects in analog black holes. The inclusion and study of astrophysical environments around black holes is currently one of the most promising directions in gravitational wave astrophysics. 
It would be very interesting if a similar experiment to the one in Ref. \cite{Svancara:2023yrf} could be extended to represent a hollow-core vortex with circulation and an acoustic horizon, with either impurities in the exterior of the horizon that do not destabilize the vortex, or nonlinear dispersive interactions of a single superfluid that discontinuously alters the velocity profile of the vortex in the radial direction. Then, the system would display an analog of a BH surrounded by an accretion disk \cite{Yuan:2025fde}, black hole hair \cite{Vlachos:2021weq,Chatzifotis:2021pak}, and even dark matter halos around supermassive black holes in galactic cores \cite{Cardoso:2021wlq,Destounis:2022obl}. The information extracted from such realistic analogs can be of significant importance for further understanding strong-field gravity and the behavior of matter around black holes \cite{Mollicone:2024lxy,Pezzella:2024tkf,Rosato:2024arw,Eleni:2024fgs,Destounis:2023gpw,Destounis:2023khj,Destounis:2023gpw,Cardoso:2022whc,Destounis:2021rko,Destounis:2021mqv}.  Finally, the spatial profiles of the eigensolutions found here can also provide valuable information for future experiments that include dissipation.

%%%%%%%%%%%%%%%%%%%%%%%%
\begin{acknowledgments}
This study was financed in part by the Conselho Nacional de Desenvolvimento Científico e Tecnológico -- Brasil (CNPq) -- Research Project No. 440846/2023-4 and Research Fellowship No. 201221/2024-1. H.S.V. is partially supported by the Alexander von Humboldt-Stiftung/Foundation (Grant No. 1209836). Funded by the Federal Ministry of Education and Research (BMBF) and the Baden-W\"{u}rttemberg Ministry of Science as part of the Excellence Strategy of the German Federal and State Governments -- Reference No. 1.-31.3.2/0086017037.
K.D. acknowledges financial support provided by FCT – Fundação para a Ciência e a Tecnologia, I.P., under the Scientific Employment Stimulus – Individual Call – Grant No. 2023.07417.CEECIND.
This project has received funding from the European
Union’s Horizon-MSCA-2022 research and innovation programme ``Einstein Waves'' under grant agreement No. 101131233.
\end{acknowledgments}
%%%%%%%%%%%%%%%%%%%%%%

\appendix

\section{Quasibound states with the VBK approach}\label{AppA}

Let us define a new radial coordinate $x$, as
\begin{equation}
x=1-\frac{r_{h}}{r},
\label{eq:radial_coordinate_UABHc}
\end{equation}
such that the three original singularities $(0,r_{h},\infty)$ are moved to the points $(-\infty,0,1)$. Next, we perform an $s$-homotopic transformation of the dependent variable $R(x) \mapsto y(x)$, such that

\begin{equation}\label{eq:dependent_variable_UABHc}
R(x)=x^{A_{0}}(x-1)^{A_{1}}\mbox{e}^{A_{2}x}y(x),
\end{equation}
where the coefficients $A_{0}$, $A_{1}$, and $A_{2}$ are given by
\begin{eqnarray}
A_{0}	& = & \pm\frac{i(\omega-m\Omega_h)}{2a}\label{eq:A0_UABHc}\\
A_{1}	& = & \pm\sqrt{1- \frac{\omega^{2}}{4a^2}},\label{eq:A1_UABHc}\\
A_{2}	& = & \pm\frac{im\Omega_h}{2a}.\label{eq:A2_UABHc}
\end{eqnarray}
By substituting Eqs.~(\ref{eq:radial_coordinate_UABHc})-(\ref{eq:A2_UABHc}) into Eq.~(\ref{eq:radial_equation_UABHc}), we get
\begin{align}
\nonumber
y''(x)&+\biggl(2A_{2}+\frac{1+2A_{0}}{x}+\frac{1+2A_{1}}{x-1}\biggr)y'(x)\\
&+\biggl(\frac{A_{3}}{x}+\frac{A_{4}}{x-1}\biggr)y(x)=0,
\label{eq:CHE_UABHc}
\end{align}
where the coefficients $A_{3}$ and $A_{4}$ are given by
\begin{align}
A_{3}&=\left(\frac{\omega^2-m^2\Omega_h^2}{2 a^2}\right)-2A_0(A_1 - A_2)\nonumber\\
&- (A_0+A_1-A_2)   
-\left(1 +\frac{\lambda}{2 a r_h}\right),
\label{eq:A3_UABHc4}\\
A_{4}&= -\frac{\omega^2}{2a^2} +2A_1\left(A_0+A_2\right)\nonumber\\ &+\left(A_0+A_1+A_2\right) +\left(1 +\frac{\lambda}{2 a r_h}\right).
\label{eq:A4_UABHc}
\end{align}
The radial equation (\ref{eq:CHE_UABHc}) has the form of a confluent Heun equation \cite{ronveaux1995heun,Minucci:2024qrn}, which is given by
\begin{equation}
y''(x)+\biggl(\alpha+\frac{1+\beta}{x}+\frac{1+\gamma}{x-1}\biggr)y'(x)+\biggl(\frac{\xi}{x}+\frac{\zeta}{x-1}\biggr)y(x)=0,
\label{eq:CHE_canonical}
\end{equation}
where $y(x)=\mbox{HeunC}(\alpha,\beta,\gamma,\delta,\eta;x)$ are the confluent Heun functions, with the parameters $\alpha$, $\beta$, $\gamma$, $\delta$, and $\eta$ related to $\xi$ and $\zeta$ by the following expressions
\begin{eqnarray}
\xi& = & \frac{1}{2}(\alpha-\beta-\gamma+\alpha\beta-\beta\gamma)-\eta,\label{eq:xi_CHE}\\
\zeta& = & \frac{1}{2}(\alpha+\beta+\gamma+\alpha\gamma+\beta\gamma)+\delta+\eta.\label{eq:zeta_CHE}
\end{eqnarray}
%%%%%%
A general analytic solution for the radial part of the covariant massless Klein-Gordon equation in the effective geometry Eq. \eqref{eq:metric_UABHc} can be written as
\begin{equation}
R_{j}(x)=x^{A_{0}}(x-1)^{A_{1}}\mbox{e}^{A_{2}x}\{C_{1,j}y_{1,j}(x)+C_{2,j}y_{2,j}(x)\},
\label{eq:solution_radial_UABHc}
\end{equation}
where $C_{1,j}$ and $C_{2,j}$ are constants to be determined, and $j=\{0,1,\infty\}$ labels the solutions at each singular point. Thus, the pair of linearly independent solutions at $x=0$ ($r=r_{h}$) is given by
\begin{eqnarray}
y_{1,0}(x) & = & \mbox{HeunC}(\alpha,\beta,\gamma,\delta,\eta;x),\label{eq:y10}\\
y_{2,0}(x) & = & x^{-\beta}\mbox{HeunC}(\alpha,-\beta,\gamma,\delta,\eta;x).\label{eq:y20}
\end{eqnarray}
Similarly, the pair of linearly independent solutions at $x=1$ ($r=+\infty$) is given by
\begin{eqnarray}
y_{1,1}(x) & = & \mbox{HeunC}(-\alpha,\gamma,\beta,\tilde{\delta},\tilde{\eta};1-x),\label{eq:y11}\\
y_{2,1}(x) & = & (1-x)^{-\gamma}\mbox{HeunC}(-\alpha,-\gamma,\beta,\tilde{\delta},\tilde{\eta};1-x),\label{eq:y21}
\end{eqnarray}
where $\tilde{\delta}$ and $\tilde{\eta}$ are obtained from Eqs.~(\ref{eq:xi_CHE}) and (\ref{eq:zeta_CHE}) by setting $\xi \rightarrow -\zeta$ and $\zeta \rightarrow -\xi$, respectively. Finally, the two linearly independent solutions of the confluent Heun equation at $|x|=\infty$ ($r=0$) can be expanded (in a sector) in the following asymptotic series
\begin{eqnarray}
y_{1,\infty}(x) & = & x^{-(\frac{\beta+\gamma+2}{2}+\frac{\delta}{\alpha})},\label{eq:y1i}\\
y_{2,\infty}(x) & = & x^{-(\frac{\beta+\gamma+2}{2}-\frac{\delta}{\alpha})}\mbox{e}^{- \alpha x}.\label{eq:y2i}
\end{eqnarray}
In these solutions, the parameters $\alpha$, $\beta$, $\gamma$, $\delta$, and $\eta$ are given by
\begin{eqnarray}
    \alpha & = & \frac{i}{a}m\Omega_h\\
    \beta & = & -\frac{i}{a}(\omega-m\Omega_{h}),\\
    \gamma & = & \frac{\sqrt{4a^{2}-\omega^{2}}}{a},\\
    \delta & = & -\frac{m^2\Omega_h^2}{2a^{2}},\\
    \eta & = & \frac{1}{2a^2}\biggl( m^2\Omega_h^2 -\omega^2+2a^2+\frac{a\lambda}{r_h}\biggr).
\label{eq:parameters_UABHc}
\end{eqnarray}

Next, we will impose specific QBS boundary conditions on the general analytic solution to compute the QBS spectrum. The first boundary condition, related to QBSs, requires that the radial solution must be purely ingoing at the acoustic event horizon (as in QNMs), and hence this is satisfied by choosing the signs ($-,+,+$) on the coefficients ($A_{0},A_{1},A_{2}$) and imposing $C_{2,0}=0$ in Eq.~(\ref{eq:solution_radial_UABHc}), such that
\begin{equation}
\lim_{r \rightarrow r_{h}} y_{1,0}(r) \sim C_{1,0}\ (r-r_{h})^{-\frac{i}{2\kappa_{h}}(\omega-\omega_{c,h})},
\label{eq:1st_condition_UABHc}
\end{equation}
where $\omega_{c,h}=\omega_{c}(r_{h})=m\Omega_{h}$.

The second boundary condition, related to QBSs, requires that the radial solution must vanish at asymptotic infinity (in contrast to QNMs that are purely outgoing there). This is satisfied when $\mbox{Re}(\sigma) > 0$, where the coefficient $\sigma$ is given by
\begin{equation}
\sigma=1+\sqrt{1-\frac{\omega^2}{4a^2}},
\label{eq:sigma_UABHc}
\end{equation}
such that
\begin{equation}
\lim_{r \rightarrow \infty} y_{1,1}(r) \sim C_{1,1}\ \frac{1}{r^{\sigma}},
\label{eq:2nd_condition_UABHc}
\end{equation}
where all the remaining constants are included in $C_{1,1}$.

According to the VBK method, these two asymptotic solutions are matched on their overlap region of validity when the confluent Heun functions become polynomials, which occurs if and only if they satisfy the following two conditions, i.e., \cite{Fiziev:2010}
\begin{eqnarray}
\frac{\delta}{\alpha}+\frac{\beta+\gamma+2}{2}+n & = & 0,\label{eq:delta-condition}\\
\Delta_{n+1}(\xi) & = & 0,\label{eq:Delta-condition}
\end{eqnarray}
where $n(=0,1,2,\ldots)$ is again the overtone number.

The first polynomial condition, given by Eq.~(\ref{eq:delta-condition}), is called as the $\delta$-condition, whose solutions are resonant frequencies; the exact spectrum of QBSs:
\begin{align}\nonumber
\omega_{mn}&=\left(\frac{Cm}{r_h^2}\right)\frac{C^2 m^2+a^2 (n^2+2 n+2) r_{h}^4}{C^2 m^2+a^2 (n+1)^2 r_{h}^4}\\
&-i a (n+1)\frac{ C^2 m^2 + a^2 n(n+2) r_{h}^4}{C^2 m^2+a^2 (n+1)^2 r_{h}^4}\, .
\label{eq:omega_UABHc}
\end{align}
These QBS frequency eigenvalues do not depend on the separation constant $\lambda$, since the parameter $\eta$ does not appear in the $\delta$-condition.

The second polynomial condition, given by Eq.~(\ref{eq:Delta-condition}), is called as the $\Delta$-condition, whose solutions are eigenvalues of the confluent Heun functions; in this case, they are the complete set of (complex) values for the separation constant $\lambda$, from which it is easy to obtain the corresponding values for the (complex) angular momentum $\nu$. However, the final expressions for $\lambda$ and $\nu$ are quite intricate, so we omit them here. Finally, to validate the VBK approach is to check the coefficient values $\sigma$ by substituting Eq.~(\ref{eq:omega_UABHc}) into Eq.~(\ref{eq:sigma_UABHc}); as expected they satisfy the requirement $\mbox{Re}(\sigma) > 0$.
%%%%%%%%%%%%%%%%%%%%%
\bibliography{biblio}
%%%%%%%%%%%%%%%%%%%%%

\end{document}